\documentclass[doublecol]{epl2}

\usepackage{ifpdf}
\usepackage{hyperref}
\usepackage{bm}
\usepackage{amsfonts}
\usepackage{amssymb}
\usepackage{graphicx}
\newcommand{\be}{\begin{equation}}
\newcommand{\ee}{\end{equation}}
\newcommand{\ben}{\begin{eqnarray}}
\newcommand{\een}{\end{eqnarray}}
\newcommand{\bes}{\begin{subequations}}
\newcommand{\ees}{\end{subequations}}
\newcommand{\wt}{\widetilde}

\title{Deformation method for generalized Abelian Higgs-Chern-Simons
models}

\shorttitle{Deformation method for Abelian HCS models}

\author{L. Losano\inst{1,2,3} \and J. M. C. Malbouisson\inst{1,4} \and
D. Rubiera-Garcia\inst{5} \and C. dos Santos\inst{1}}
\shortauthor{L. Losano \etal}

\institute{
  \inst{1} Centro de F\'{\i}sica e Departamento de F\'{\i}sica
e Astronomia, Faculdade de Ci\^{e}ncias da Universidade do Porto,
4169-007 Porto, Portugal\\
  \inst{2} Departamento de F\'isica, Universidade Federal da
Para\'\i ba, 58051-900 Jo\~ao Pessoa, PB, Brazil\\
  \inst{3} Departamento de F\'isica, Universidade Federal de
Campina Grande, 58109-970 Campina Grande, PB, Brazil\\
  \inst{4} Instituto de F\'{\i}sica,
Universidade Federal da Bahia, 40210-340 Salvador, BA, Brazil\\
  \inst{5} Departamento de F\'{\i}sica, Universidad de Oviedo,
33007 Oviedo, Asturias, Spain }


\newcommand{\bh }{\begin{displaymath}}
\newcommand{\eh }{\end{displaymath}}

\abstract{ We present an extension of the deformation method applied
to self-dual solutions of generalized Abelian Higgs-Chern-Simons
models. Starting from a model defined by a potential $V(\vert \phi
\vert)$ and a non-canonical kinetic term $\omega(\vert \phi \vert)
\vert D_{\mu}\phi \vert^2$ whose analytical domain wall solutions
are known, we show that this method allows to obtain an uncountable
number of new analytical solutions of new models defined by other
functions $\widetilde{V}$ and $\widetilde{\omega}$. We present some
examples of deformation functions leading to new families of models
and their associated analytic solutions. }

\begin{document}

\maketitle

\section{Introduction}

Topological defects play important role in several important areas
such as high energy physics~\cite{solitons},
cosmology~\cite{Cosmology} and condensed matter
physics~\cite{condensed}. Such defects emerge as classical solutions
of nonlinear field theories which possess degenerated vacua. Typical
examples are domain walls described by kink solutions of the
$\phi^4$ model, Ginzburg-Landau vortices and monopoles.

Usually, domain walls are solutions connecting two distinct vacua of
scalar field theories in one-space dimension, or in their insertions
in higher dimensions, while vortices emerge as solutions of models
that couple charged matter fields with gauge fields living in a (at
least) 3-dimensional space-time, and monopoles lie in a 4-D
space-time.

In a (2+1)-dimensional space-time, minimal coupling between
charged-matter and gauge fields can be implemented by the
Chern-Simons (CS) action. Although the CS field can not be conceived
as a free field, its coupling with matter fields imposes constraints
in the dynamics which have very relevant consequences, in both,
classical and quantum theories, with either relativistic or
non-relativistic kinetics. In the non-relativistic (NR) framework,
particles coupled through the CS field carry both electric charge
and magnetic flux, and possess fractional statistics~\cite{Lerda}.
Additionally, the NR scalar CS model constitutes a seminal example
of a Galilean-invariant gauge-field theory~\cite{Hagen}. Also, for a
critical strength of a quartic self-interaction of the scalar field,
which restores the scale invariance~\cite{JackiwPi}, this model
provides a field-theoretical description of the Aharonov-Bohm (AB)
scattering~\cite{BergmanLozano}; considering the Lorentz covariant
field theory, relativistic corrections to the AB scattering are
obtained~\cite{GMS}.

Self-dual soliton solutions have been found in the relativistic,
${\rm U}(1)$-invariant, Abelian, Higgs-Chern-Simons (HCS)
gauge-theory where the symmetry-breaking potential of the Higgs
field is $U(\varphi) \sim
|\varphi|^2(|\varphi|^2-v^2)^2$~\cite{HongJackiw}; vortex and
domain-wall solutions have been obtained for this model~\cite{JLW}.
This model was generalized by considering a non-canonical kinetic
term for the complex scalar field, ${\cal
W}(|\varphi|)|D_{\mu}\varphi|^2$, providing self-dual
vortex~\cite{Bazeia10} and domain-wall~\cite{Santos10} solutions.
Models with noncanonical kinetic terms (k-fields) find also
applications in strong-interaction physics~\cite{strong} and in
cosmology~\cite{CosmologyK}.

Due to the nonlinearity, there is no general integration method to
solve analytically the equations of motion of non-linear field
theories; only for a small set of models, solutions of the equations
of motion can be directly determined. However, for scalar fields in
(1+1)-dimensions, starting from a nonlinear model with known
solutions, infinitely many new models and their corresponding static
solutions can be found using the deformation
method~\cite{DeforMethod}. This method works as follows. Choosing a
deformation function $f(\phi)$, the model defined by the deformed
potential $\wt{V}(\phi) = V[(f(\phi))]/[f^{\prime}(\phi)]^2$, where
$f^{\prime}$ means the derivative of $f$, possesses static solutions
given by $\wt{\phi}(x) = f^{-1}(\phi(x))$, where $\phi(x)$ is a
solution of the static equation of motion of the original model with
potential $V(\phi)$. This procedure has been applied to generate
defect solutions of many models having polynomial
interactions~\cite{Deformation2} and new families of sine-Gordon and
multi-sine-Gordon models~\cite{Deformation3}. Also, an orbit-based
extension of this method has been applied to models involving two
interacting scalar fields~\cite{Afonso07}.

The purpose of this Letter is to extend the deformation method to
gauge-field models considering specifically the Abelian HCS theory,
focusing particularly on the Jackiw-Lee-Weinberg (JLK) domain-wall
solution~\cite{JLW}. In Section II, we present the generalized
Abelian HCS models and write down the first-order equations obeyed
by the Bogomol'nyi-Prasad-Sommerfeld (BPS)~\cite{BPS} domain-wall
solutions. In Section III, the deformation method is extended to
domain-wall solutions of generalized Abelian HCS models and some
examples are given, illustrating the power of the procedure in
generating new models with their static solutions. Finally, some
remarks are made.

\section{BPS domain walls in the generalized Abelian HCS model}

We consider the generalized $(2+1)$-dimensional Abelian HCS model
defined by the Lagrangian density~\cite{Bazeia10}
\begin{equation}
{\cal L_S} = {\cal W}(|\varphi|)|D_{\mu}\varphi|^2-U(|\varphi|) +
\frac{\kappa}{4}\epsilon^{\alpha\beta\gamma}{\cal A}_{\alpha}{\cal
F}_{\beta \gamma} , \label{lagrangian}
\end{equation}
where $\varphi$ is the complex Higgs field, $D_{\mu}=\partial_{\mu}
+ ie {\cal A}_{\mu}$ is the covariant derivative and ${\cal
F}_{\mu\nu}=\partial_{\mu}{\cal A}_{\nu}-\partial_{\nu}{\cal
A}_{\mu}$ is the field strength tensor of the gauge potential ${\cal
A}_{\mu}$. The self-interaction potential, $U(\vert \varphi \vert)$,
is assumed to implement a symmetry-breaking mechanism and the
non-canonicity of the kinetic term is engendered by the function
${\cal W}(\vert \varphi \vert)$; taking ${\cal W}\equiv 1$, one
recovers the standard Abelian HCS model. Note that, in the CS term,
$\epsilon^{\alpha\beta\gamma}$ is the fully antisymmetric tensor and
the electric and the magnetic CS fields are ${\cal E}^i={\cal
F}^{i0}=-{\dot {\cal A}}^i - \nabla_i {\cal A}^0$ and ${{\cal
B}}=\vec{\nabla} \times \vec{{\cal A}} = \partial_2 {\cal A}_1 -
\partial_1 {\cal A}_2$, respectively.

It is convenient to work with dimensionless quantities. In $(2+1)$
dimensions, the scalar field $\varphi$ has mass dimension equal to
$1/2$, the same one we take for the gauge field; this choice ensures
that the mass dimension of ${\cal A}_{\alpha}$ agrees with the one
obtained if a Maxwell term were added to ${\cal L_S}$. It follows
that the electric charge $e$ and the CS parameter $\kappa$ has mass
dimensions equal to $1/2$ and $1$, respectively, so that
$e^2/\kappa$ is dimensionless. We can get an additional
simplification if we absorb the parameters $e$ and $\kappa$ by
redefining space-time coordinates and fields. Thus, with $M$ being a
mass scale of the model, we define ${\bar x}^{\mu} =
M{e^2}x^{\mu}/\kappa$, $\phi={\sqrt{\kappa}}\varphi/{\sqrt{M} e}$,
$A_{\mu} = \kappa {\cal A}_{\mu}/{M e}$, $V=\kappa^2 U/{M^3 e^4}$
and $\omega = e^2{\cal W}/\kappa$; the dimensionless Lagrangian
density is then given by ${\cal L}= \kappa^2{\cal L_S}/{M^3 e^4}$
and the action becomes ${\cal S} = \frac{\kappa}{e^2}\int d^3 {\bar
x} {\cal L}$. For a simpler notation, we suppress the bar over the
space-time coordinates and use, from now on, only dimensionless
quantities.

Variation of the action leads to the equations of motion \ben \omega
D_{\mu}D^{\mu}\phi &+&
\partial_{\mu} \omega D^{\mu} \phi - \vert D_{\mu} \phi \vert^2
\frac{\partial \omega}{\partial \phi^{*}}
+\frac{\partial V}{\partial \phi^*} =0\,,\label{eq:scalar}\\
&&\frac{1}{2}\epsilon^{\alpha\beta\gamma}F_{\beta\gamma}=-J^{\alpha}\,,
\label{eq:gauge} \een where the current density, $J^{\alpha}=(\rho,
\vec{j})$, is given by \be J^{\alpha}=i\omega\left[\phi (D^{\alpha}
\phi)^* - \phi^* D^{\alpha} \phi\right]\,. \ee The time component of
eq.~(\ref{eq:gauge}) states that the magnetic field is equal to the
planar electric-charge density, $B = \rho$, which is the CS Gauss
law. Also, for static field configurations, we find \be
{B}\,=\rho=2A_0|\phi|^2\omega(|\phi|)\,,\,\,\,\,\,\,\,
{E}^a=\epsilon_{ab}j^b\,,\label{B} \ee which shows that the
electric-current density is perpendicular to the electric field.

The energy-momentum tensor is given by \ben\label{emtensor}
T_{\mu\nu} & = & \omega \left[ D_{\mu} \phi \, (D_{\nu} \phi)^* +
D_{\nu} \phi \, (D_{\mu} \phi)^* \right] \nonumber \\
& & -\, g_{\mu\nu} \left[\omega \vert D_{\alpha} \phi \vert^2
-V(\vert \phi \vert) \right] \een  from which we obtain the energy
density, $\varepsilon=T_{00}$, and the pressure components, ${\cal
P}_1=T_{11}$ and   ${\cal P}_2=T_{22}$.

We are interested in static domain-wall solutions. Firstly, note
that the complex phase of the scalar field $\phi$ can be suppressed
by a suitable gauge transformation. Then, fixing the Coulomb gauge,
we can search for solutions of the form~\cite{JLW,Santos10} \be
\phi=h(x)\,,\,\,\,\,\,\,A_\mu=\big(A_0(x), A_1=0, A_2=A(x)\big)\,,
\ee where $h(x)$ and $A(x)$ are real functions and $x$ denotes the
$x^1$-coordinate. This ansatz corresponds to domain-walls (actually
lines in the plane) parallel to the $x^2$-axis.

In this case, the static equations of motion reduces to \ben \left[2
\omega h^\prime\right]^\prime & = & 2 h \omega
\left(A^2-A_0^2\right) + \frac{dV}{dh}\, ,\label{heqm}\\
A_0^\prime & = & - 2 \omega h^2 A\label{Aeqm}\,, \een and the Gauss
law \be\label{gauss} A^\prime = -2 \omega h^2 A_0\,, \ee where the
prime denotes derivation with respect to $x$. From eqs.~(\ref{Aeqm})
and (\ref{gauss}) we infer that $A_0 A_0^\prime = A A^\prime$, so
that time and space components of the gauge filed are constrained by
\be A_0^2=A^2-C\,, \ee where $C$ is a real constant. Also,
consistency with eq.~(\ref{heqm}) imposes a relation between the
function $\omega(h)$ and the potential $V(h)$ expressed as \be
\frac{d}{dh}\Bigg[\frac{\sqrt{V/\omega}}{h}\Bigg]=-2\omega h\,. \ee

Now, the stability condition ${\cal P}_1={\cal P}_2=0$ leads to the
first-order equations~\cite{Vega1976}
\begin{eqnarray}
h'&=&\pm hA\label{hdef}\\
A'&=&-2\omega h^2 A_0\label{Adef}
\end{eqnarray}
with
\begin{equation} \label{eq:A0}
V=h^2 \omega A_0^2\,.
\end{equation}
For $h\geq 0$ and $A\geq 0$, the signal $+$ ($-$) in
eq.~(\ref{hdef}) corresponds to the kink (anti-kink) like solution
for the Higgs field, $h^{(+)}$ ($h^{(-)}$). Note that, the
first-order equations (\ref{hdef}) and (\ref{Adef}) solve the
equations of motions (\ref{eq:scalar}) and (\ref{eq:gauge}).

The static solutions are physically characterized by their charge
and energy. Now, returning to eq.~(\ref{emtensor}), for non-negative
$V(h)$ and $\omega(h)$,   the energy of static solutions can be
rewritten in the form \ben E & = &\int_{-\infty}^{\infty}dx\,T_{00}
 \nonumber\\ &=& \int_{-\infty}^{\infty}dx\,
(V+\omega h'+2\omega h^2 A_0^2+C\omega h^2) \nonumber\\
&=&\int_{-\infty}^{\infty}dx\,\Big[\left(\sqrt{V}\pm\sqrt{\omega}hA_0^2
\right)^2+\left(\sqrt{\omega}h' \pm\sqrt{\omega}hA\right)^2\nonumber\\
&&+\left(\sqrt{-A_0A'}\pm\sqrt{2\omega}hA_0\right)^2\Big]\nonumber\\
&&+\int_{-\infty}^{\infty}dx\,\Big(2\sqrt{\omega V}hA_0\pm2\omega hh'
A \nonumber\\
&&\pm 2\sqrt{-2\omega A_0A'}hA_0 + A_0A'-2\omega h^2A_0^2\Big)\,,
\een which is minimized if eqs.~(\ref{hdef}), (\ref{Adef}), and
(\ref{eq:A0}) are obeyed, resulting in \be\label{E} E=
\int_{-\infty}^{\infty}dx\,(4V)
=\left|\left(A^2(-\infty)-A^2(+\infty)\right)\right |\,, \ee for
$C=0$.  In this case  $A_0^2=A^2$, so the system of first-order
equations decouples and is solved simply by (\ref{hdef}) with
\be\label{Ah} A(h) = -2\int {\omega h}\,dh + c \, , \ee where $c$ is
an integration constant suitable to the boundary conditions required
for the gauge field. And, from (\ref{B})  and (\ref{emtensor}), the
electric charge, $Q$,  and Noether charge, $P$,  are given by
\be\label{Q} Q=\int^\infty_{-\infty}dx
\rho(x)=A(-\infty)-A(+\infty)\,, \ee \be\label{P}
P=\int^\infty_{-\infty} dx\,
T_{02}=\frac12\left[A^2(-\infty)-A^2(+\infty)\right]\,, \ee which
are both conserved due to the $U(1)$ symmetry and the translational
invariance along $x^2$-direction, respectively.

This shows that, for $h$ in a range such that $\omega(h)\geq 0$ and
$V(h)\geq 0$, the BPS solutions of the first-order eqs.~(\ref{hdef})
and (\ref{Adef}), with (\ref{eq:A0}), indeed correspond to solutions
of minimum energy and their energy and charge can be calculated
knowing only the asymptotic behavior of the gauge field.
Correspondingly, the Higgs field, for both kink and anti-kink
solutions, connects two consecutive vacua of the potential, while a
lump-like solution starts and terminates on the same vacuum when
$x\rightarrow\pm\infty$.


\subsection{Standard self-dual domain walls}\label{jack}

\begin{figure}[ht]
\includegraphics[{height=02cm,width=07cm,angle=00}]{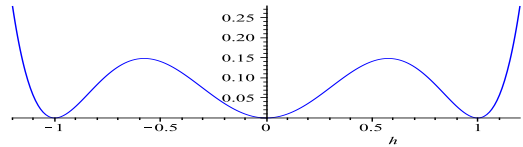}
\caption{The potential (\ref{Vjac}) as function of the Higgs
field.}\label{fig1a}
\end{figure}

\begin{figure}[ht]
\includegraphics[{height=2.5cm,width=4cm,angle=00}]{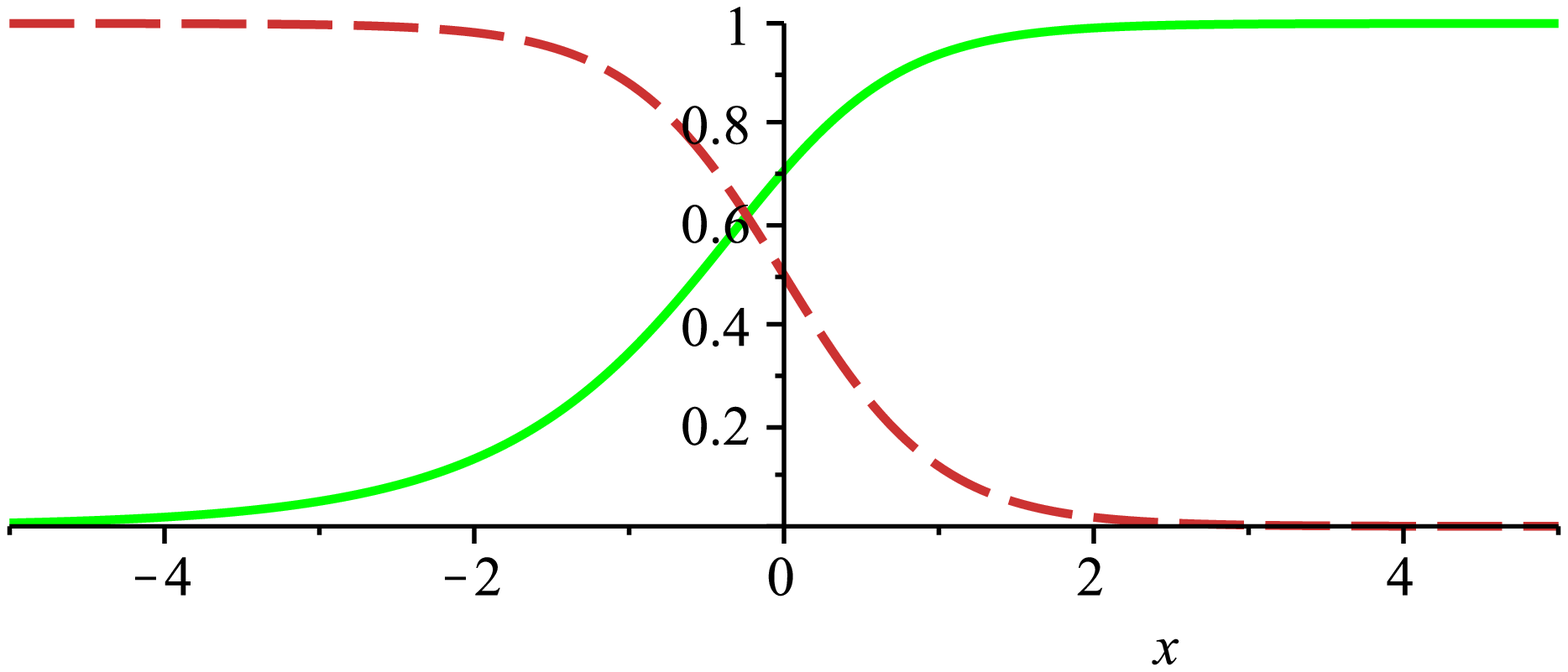}
\includegraphics[{height=2.5cm,width=4cm,angle=00}]{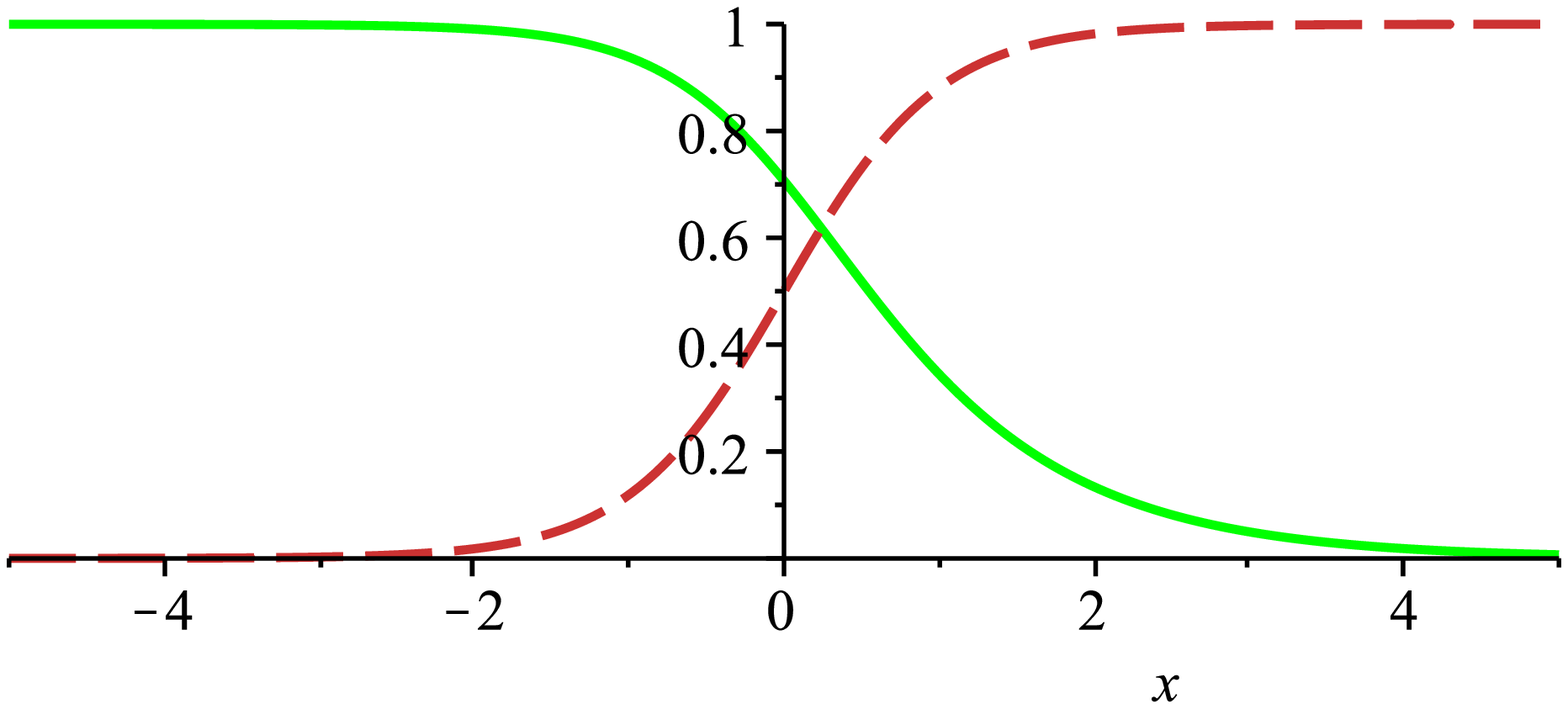}
\caption{The Higgs field  (solid line) and the gauge field (dashed
line),  $(h^{(+)}(x), A^{(-)}(x))$ from eq.~(\ref{hjac1}) on the
left, and $(h^{(-)}(x), A^{(+)}(x))$ from eq.~(\ref{hjac2}) on the
right.} \label{fig1b}
\end{figure}

\begin{figure}[ht]
\includegraphics[{height=2.5cm,width=07cm,angle=00}]{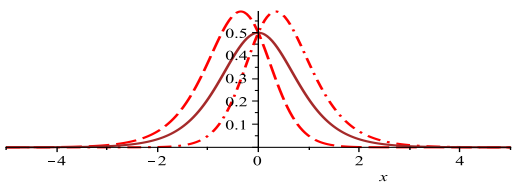}
\caption{Energy density of the solutions $(h^{(+)}(x), A^{(-)}(x))$
(dashed line) and $(h^{(-)}(x), A^{(+)}(x))$ (dashed-dotted line),
and module of electric-charge density for both solutions (solid
line).}\label{fig1d}
\end{figure}

The simplest Abelian HCS model that supports self-dual domain wall
solutions is the JLW model~\cite{JLW}, which is defined by the
Lagrangian~\ref{lagrangian} with canonical kinetic term ($\omega=1$)
and the (dimensionless) potential
\begin{equation}\label{Vjac}
V(h) = h^2(1-h^2)^2,
\end{equation}
plotted in fig.~\ref{fig1a}.
In this case, the use of eq.~(\ref{Ah}) (with $c=1$) provides the
result \be\label{Ajac} A=1-h^2, \ee which, substituting
in~(\ref{hdef}), gives the solutions \be h^{(+)}(x)
=1/{\sqrt{1+e^{-2x}}}\,,\,A^{(-)}(x)=1/({1+e^{2x}}),\label{hjac1}
\ee and \be h^{(-)}(x)
=1/{\sqrt{1+e^{2x}}}\,,\,A^{(+)}(x)=1/({1+e^{-2x}}),\label{hjac2}
\ee which are displayed in fig.~\ref{fig1b}. We see that the scalar
field, in both cases, interpolates between the symmetric and the
asymmetric vacua.
fig.~\ref{fig1d}shows the energy and electric-charge densities for
both wall solutions. We find that the spatial distribution of the
electric charge is symmetric around the origin, while for the energy
the axis of symmetry are displaced from the origin.
And, from eqs.~(\ref{E}), (\ref{Q}) and (\ref{P}), for the solutions
$(h^{(+)}, A^{(-)} )$ and $(h^{(-)}, A^{(+)} )$, we have the charges
$Q=1$, $P=1/2$, and $Q=-1$, $P=-1/2$, respectively, and the same
energy, $E=1$.


\section{The deformation method}

Let us now develop the deformation method for generalized Abelian
HCS models following the spirit of the procedure introduced for
scalar fields~\cite{DeforMethod}. As we shall show, by deforming
simultaneously the Higgs and the CS fields, we are able to construct
many new generalized HCS models and their static domain-wall
solutions. The original and the deformed models are mapped into each
other through the deformation function.

Denote by $\tilde{\phi}(x)$ and $\wt{A}(x)$ new fields whose
dynamics is governed by the (dimensionless) Lagrangian density
\begin{equation} \label{lagrangian2}
\widetilde{{\cal L}} =
\widetilde{\omega}(|\tilde{\phi}|)|D_{\mu}\tilde{\phi}|^2 -
\widetilde{V}(|\tilde{\phi}|) +
\frac{1}{4}\epsilon^{\alpha\beta\gamma}\widetilde{A}_{\alpha}
\widetilde{F}_{\beta \gamma}\,,
\end{equation}
where $\widetilde{V}(\vert \tilde{\phi} \vert)$ and
$\widetilde{\omega}(\vert \tilde{\phi}\vert)$ are new functions
specifying this model. As in sec.~II, we assume that the self-dual
BPS domain-wall solutions of this model take the form \be
\tilde\phi=\tilde h(x)\,,\,\,\,\,\,\,\widetilde
A_{\mu}=\big(\widetilde A_0(x), \widetilde A_1=0, \widetilde
A_2=\widetilde A(x)\big)\,, \ee and satisfy the first-order
equations of motion \ben
\tilde{h}'&=&\pm\tilde {h}\widetilde{A}\,, \label{dhtilde} \\
\widetilde{A}'& = &-2\widetilde{\omega} \tilde{h}^2\widetilde{A}_0
\label{dAtilde}\,, \een where $\tilde{h}' \equiv{d\tilde{h}}/{dx}$
and $\widetilde{A}' \equiv {d\widetilde{A}}/{dx}$, with the
constraints $\wt{V}=h^2 \widetilde\omega \widetilde{A}_0^2$ and
$\wt{A}_0^2=\widetilde{A}^2$.

Now, introduce the deformation function $f$ such that the Higgs
fields of the two models are mapped into each other,
$h=f(\tilde{h})$, which is assumed to be invertible (in a prescribed
domain of definition) and differentiable. Also, consider that the
deformed CS-gauge field is obtained from $A$ by the prescription
\begin{equation} \label{Atilde}
\widetilde{A} (\tilde{h}) = \frac{f(\tilde{h})\,A[h \rightarrow
f(\tilde{h})]}{\tilde{h}f_{\widetilde{h}}} \,,
\end{equation}
where $f_{\widetilde{h}}={df}/{d\widetilde{h}}$. Then, it follows
from eqs.~(\ref{dhtilde}) and (\ref{dAtilde}), using
eq.~(\ref{Atilde}), that the model defined by Lagrangian density
(\ref{lagrangian2}), with the deformed function $\widetilde{\omega}$
and the deformed potential $\widetilde{V}$ given by
\be\label{wVtilde} \wt{\omega}(\tilde{h}) =
\frac{1}{2}\frac{\widetilde{A}_{\tilde{h}}}{\tilde{h}}\,,
\,\,\,\,\,\,\,\,\,\wt{V}(\tilde{h})=\tilde h^2\wt
A^2\,\wt{\omega}(\tilde{h})\,, \ee where $\widetilde{A}_{ \tilde
h}=d\widetilde{A}/d\tilde h$, possesses static BPS solutions
\begin{equation}\label{defBPS}
\tilde{h}(x) = f^{-1}[h(x)]\, , \,\,\,\,\,\,\, \widetilde{A}(x) =
\widetilde{A}\big( f^{-1}[h(x)]\big) \, ,
\end{equation}
where $h(x)$ is a static solution of the original
model~(\ref{lagrangian}).

It should be noted that all the considerations and relations
presented in sec.~II, relative to energy and conserved charges, are
held unchanged for the deformed system. In the following, taking as
the starting point the JLW domain-wall solutions described in
sec.~II.A, we consider some illustrative examples of the method.


\subsection{Example I}

\begin{figure}[ht]
\includegraphics[{height=03cm,width=07cm,angle=00}]{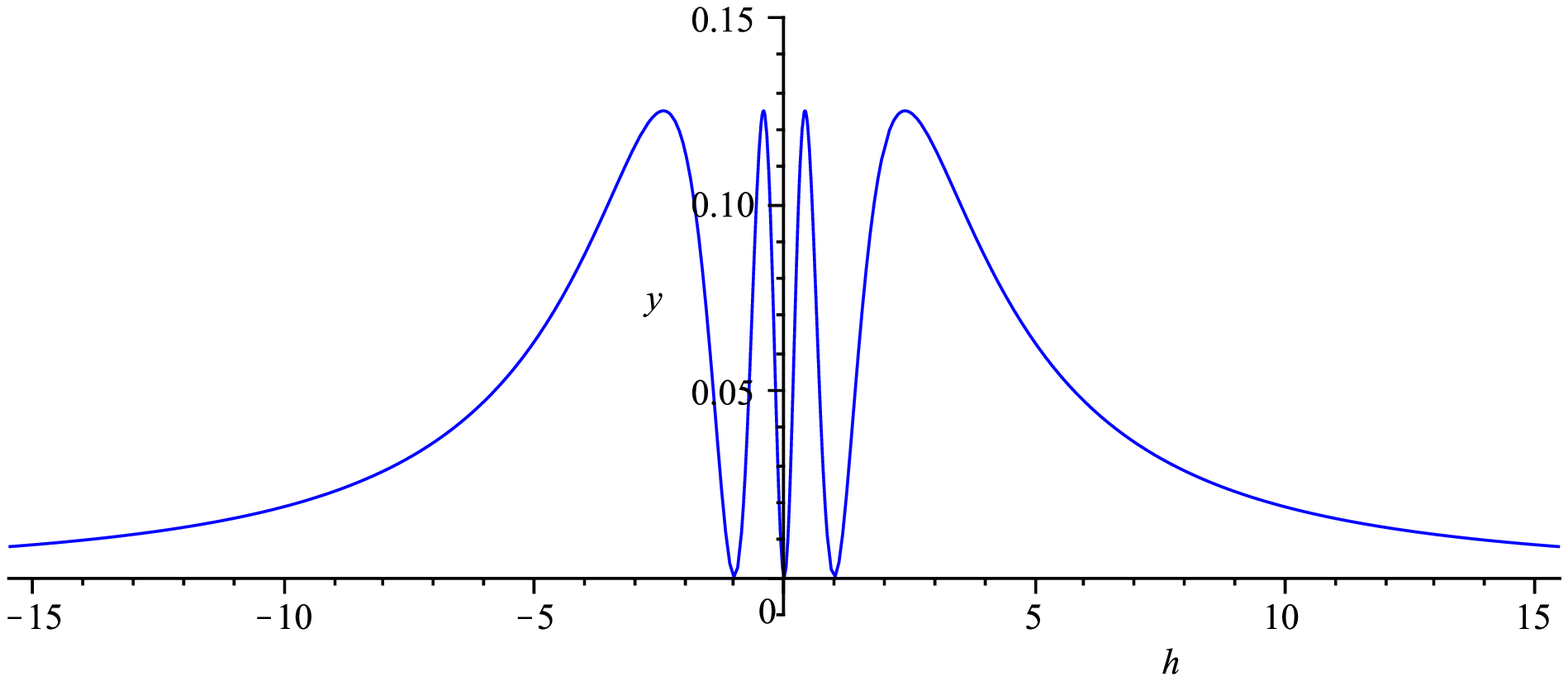}
\includegraphics[{height=02cm,width=07cm,angle=00}]{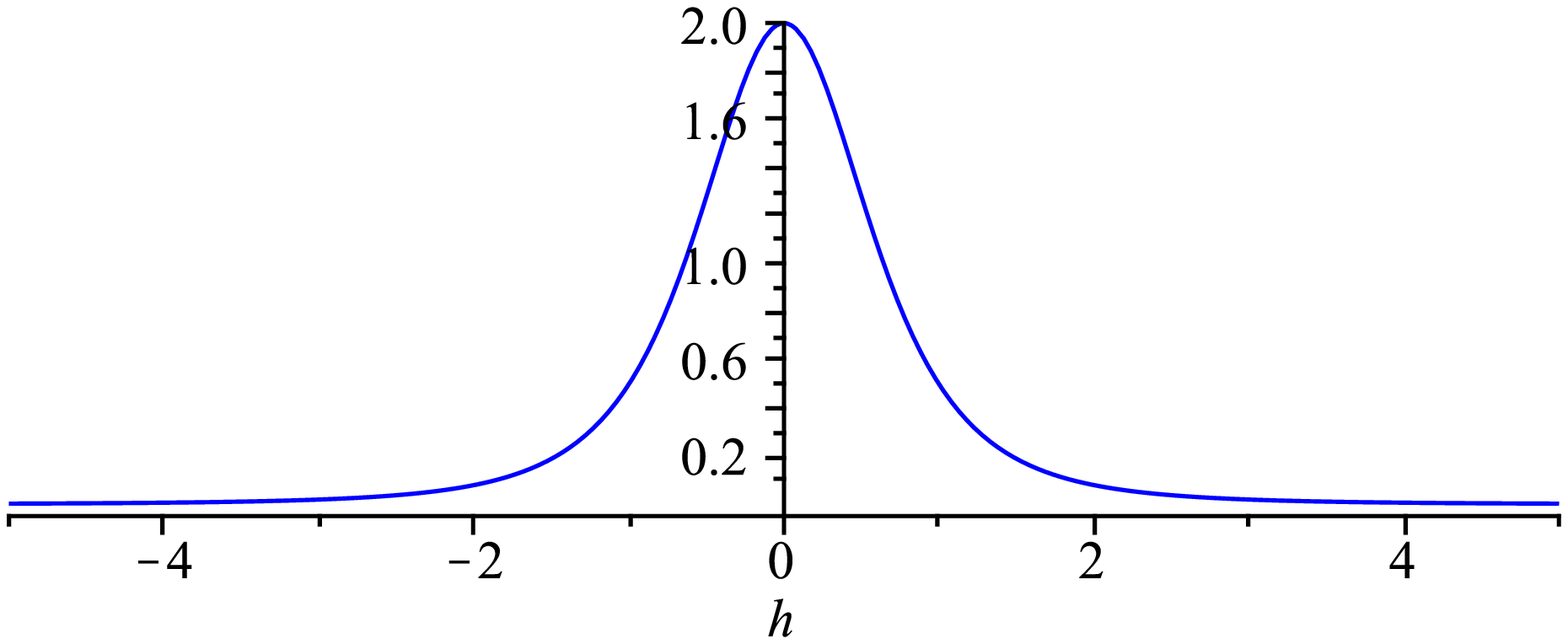}
\caption{The potential (\ref{wVdef1}) (top panel) and the
corresponding function $w$ (bottom panel), as function of the Higgs
field.}\label{fig2a}
\end{figure}

Firstly, we consider the couple of deformation function
\begin{equation} \label{def1}
f(\tilde{h})^{(\pm)}=(\pm)\frac{1-\tilde h^2}{1+\tilde h^2}\,,
\end{equation}
which, using eqs.~(\ref{Ajac}) and (\ref{Atilde}), gives
$\widetilde{A}^{(\pm)}(\tilde{h} )=f(\tilde{h})^{(\pm)}$; and, from
eq.~(\ref{wVtilde}), it follows that \be\label{wVdef1}
\tilde\omega=\frac{2}{(1+\tilde h^2)^2}\,,\,\,\,\,\wt
V=\frac{2\tilde h^2(1-\tilde h^2)^2}{(1+\tilde h^2)^4} \, . \ee
These functions, which are plotted in fig.~\ref{fig2a}, define the
generalized Abelian HCS model employed in Ref.\cite{Santos10}. Note
that, the three vacua at $\tilde h=0,1,+\infty$ establish two walls,
one between $\tilde{h}=0$ and $\tilde{h}=1$, and other between
$\tilde{h}=1$ and $\tilde{h}=+\infty$.
\begin{figure}[ht]
\includegraphics[{height=2.5cm,width=4cm,angle=00}]{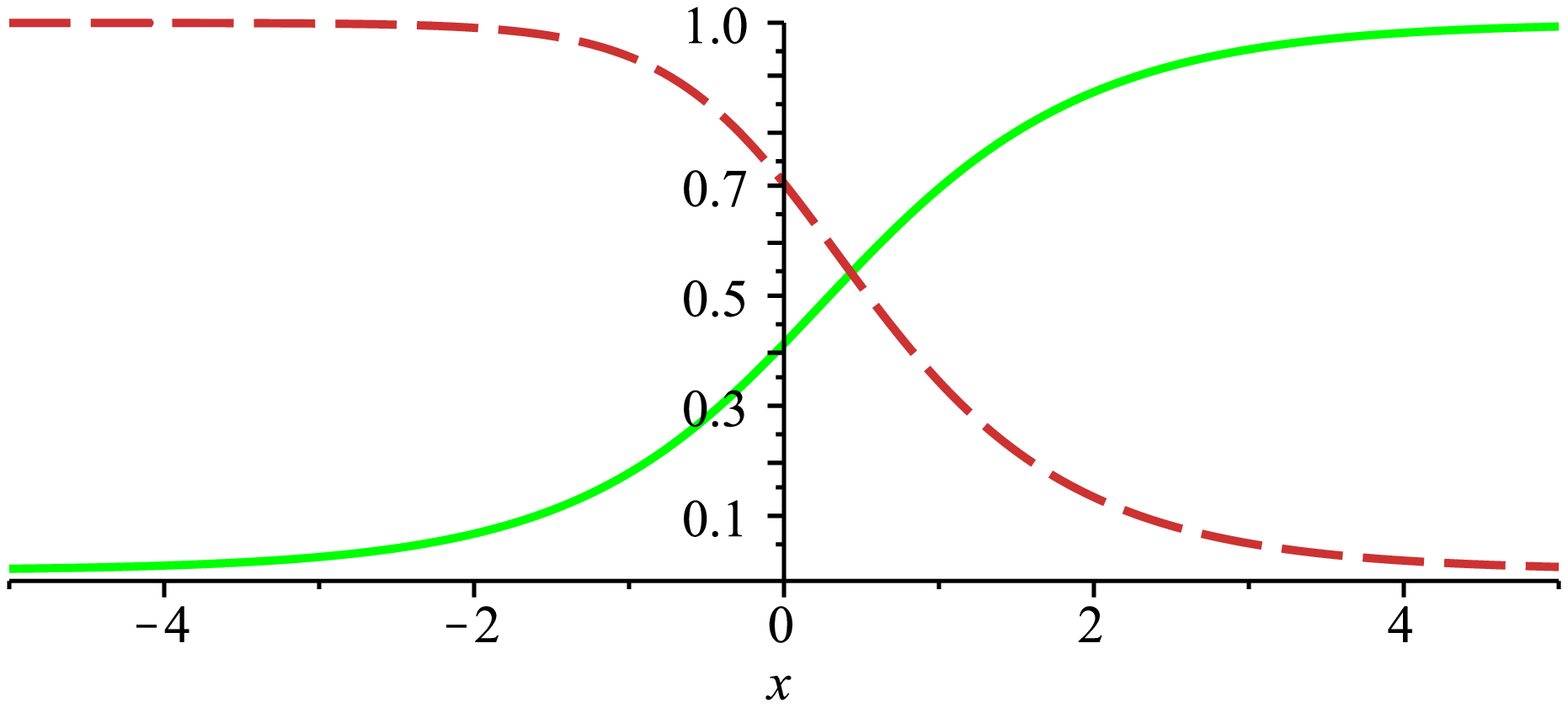}
\includegraphics[{height=2.5cm,width=4cm,angle=00}]{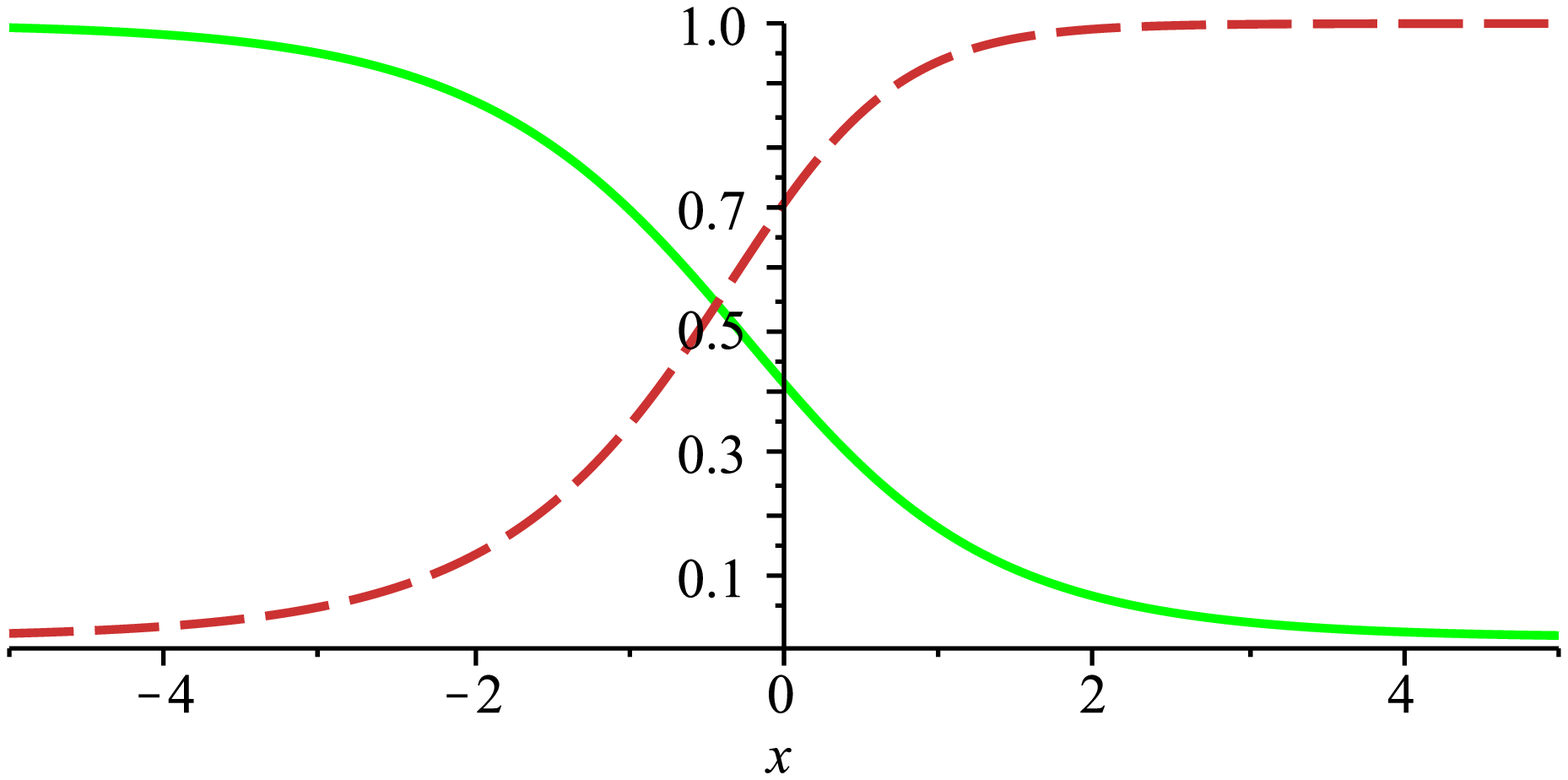}
\caption{The Higgs field  (solid line) and the gauge field (dashed
line),  $(\tilde h^{(+)}(x),\, \wt A^{(-)}(x))$ from
eq.~(\ref{h1def1a}) on the left, and  $(\tilde h^{(-)}(x),\,\wt
A^{(+)}(x))$ from eq.~(\ref{h1def1b}) on the right, for $0\leq
\tilde h\leq 1$}\label{fig2c}
\end{figure}
From the inverse of the deformation function (\ref{def1}) and
eqs.~(\ref{hjac1}) and (\ref{hjac2}), for the range $0\leq \tilde
h\leq 1$, we obtain the solutions \ben
\tilde{h}^{(+)}(x)&=&\sqrt{1+e^{-2x}}-e^{-x}\,,\,
\wt{A}^{(-)}(x)=1/{\sqrt{1+e^{2x}}}\,,\nonumber\\
&&\label{h1def1a}\\
\tilde{h}^{(-)}(x)&=&1/{\sqrt{1+2e^{2x}}}\,,\,\,
\wt{A}^{(+)}(x)=1/{({1+e^{-2x}})},\label{h1def1b} \een while for
$\tilde h\geq 1$ we have \ben
\tilde{h}^{(+)}(x)&=&{\sqrt{1+2e^{2x}}}\,,\,\,\,\,\,
\wt{A}^{(+)}(x)=1/({{1+e^{-2x}}})\,,\label{h2def1a}\\
\tilde{h}^{(-)}(x)&=&\sqrt{1+e^{-2x}}+e^{-x}\,,\,\,
\wt{A}^{(-)}(x)=1/({\sqrt{1+e^{2x}}}).\nonumber\\
&&\label{h2def1b} \een In figs.~\ref{fig2c}  and \ref{fig2e}, we
display these domain wall solutions. The walls for $0\leq \tilde
h\leq 1$ and $\tilde h\geq 1$ have the same gauge fields, but with
the asymptotic value for $x=\pm\infty$ changed. Then, for both
ranges the walls have the same energy, $E=1$, and charges $Q=1$ and
$P=1/2$, for $\wt A^{(-)}$,  and  $Q=-1$  and $P=-1/2$, for $\wt
A^{(+)}$. This makes possible to have attractive or repulsive force
between the two walls.\\
\begin{figure}[ht]
\includegraphics[{height=2.5cm,width=4cm,angle=00}]{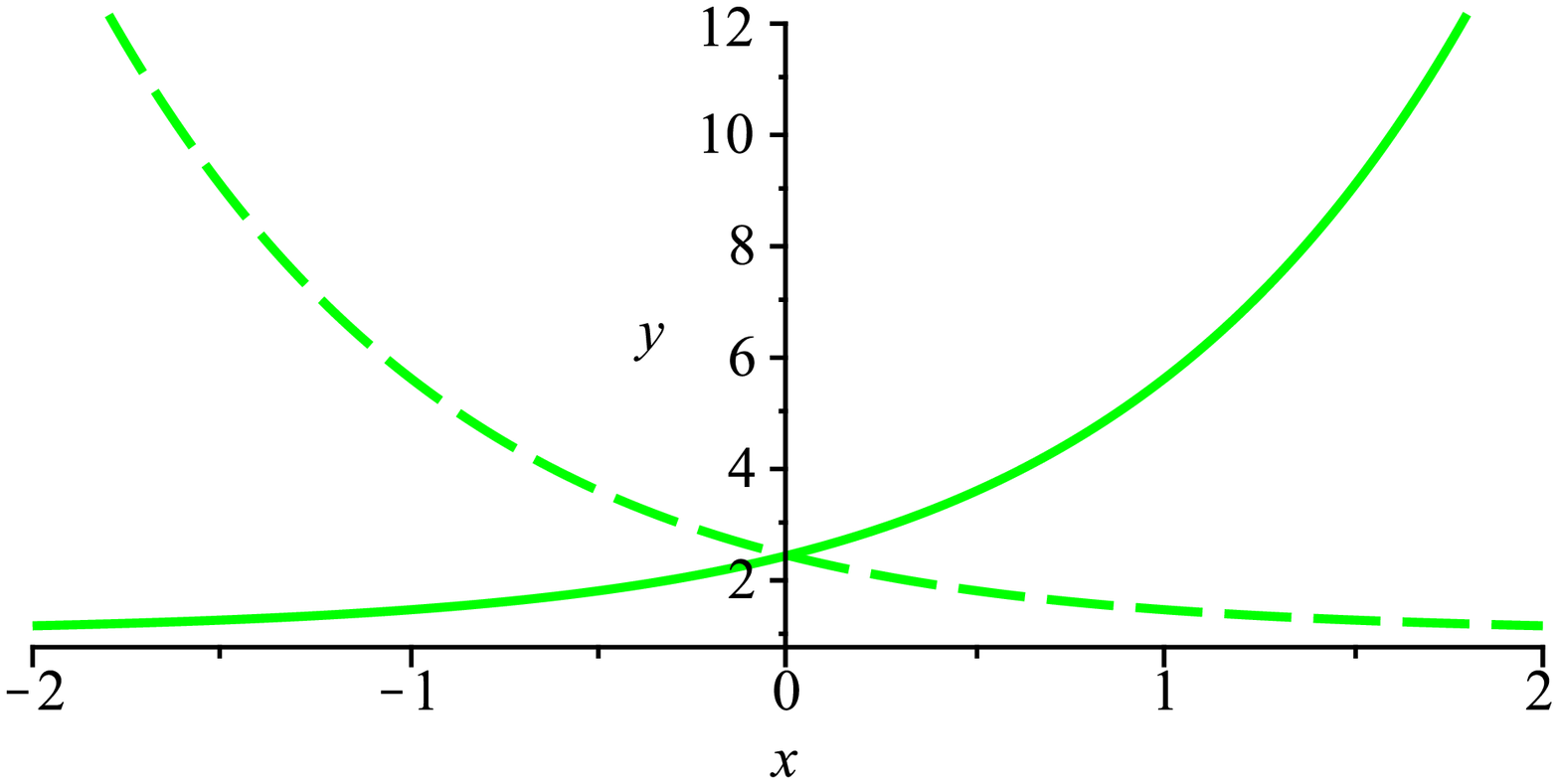}
\includegraphics[{height=2.5cm,width=4cm,angle=00}]{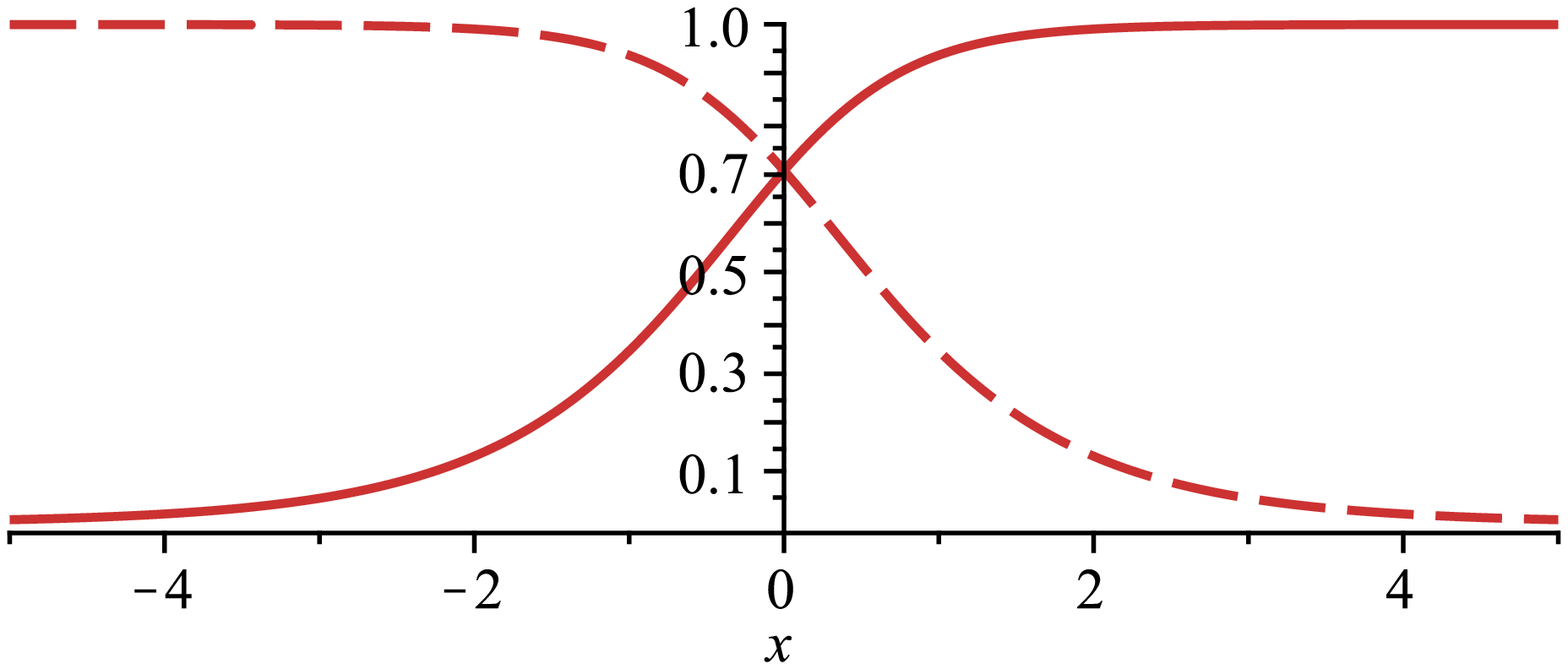}
\caption{The fields $(\tilde h^{(+)}(x),\,\wt A^{(+)}(x))$,
eq.~(\ref{h2def1a})  (solid  line),  and $(\tilde h^{(-)}(x),\,\wt
A^{(-)}(x))$, eq.~(\ref{h2def1b})  (dashed line), for $\tilde h\geq
1$.}\label{fig2e}
\end{figure}
\begin{figure}[ht]
\includegraphics[{height=2.5cm,width=7cm,angle=00}]{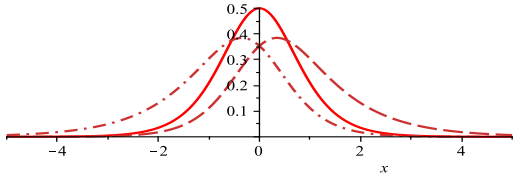}
\caption{Module of electric charge for solutions $A^{(-)}(x)$
(dashed line) and $A^{(+)}(x)$ (dashed-dotted line), and energy
density (solid line) for both walls.}\label{fig2g}
\end{figure}
In fig.~\ref{fig2g}, we display the energy and charge densities for
the two walls. The comparison with the walls of the JLW model shows
that, notwithstanding the walls have the same charges and energy,
the JLW walls have symmetric spatial distributions of energy and
charge, while here only the distribution of energy is symmetric and
all the corresponding distributions are more spread out. The model
defined by eqs.~(\ref{wVdef1}), which was obtained by deforming the
JLW model, was studied in Ref.~\cite{Santos10} but only the solution
satisfying $0\leq \tilde h\leq 1$ was considered.

The deformation function (\ref{def1}) is a particular case of the
deformation function $f(\tilde h)=\cos[\alpha\arctan(\tilde h)]$,
corresponding to $\alpha=2$; from that new family of models can be
generated for $\alpha$ integer.


\subsection{Example II}

As a second example,  consider the set of deformation functions
\cite{Deformation2} \be \label{def2}
f_\alpha(\tilde{h})]=\cos[\alpha\,\arccos(\tilde{h})] =
T_\alpha(\tilde h)\,, \ee where the integer $\alpha>2$ and
$T_{\alpha}$ is the Chebyshew polynomials of first kind. Using this
deformation in eq.~(\ref{Atilde}), with eq.~(\ref{Ajac}), we have
the gauge field \ben \wt{A}_\alpha (\tilde{h} )& = &({1-\tilde
h^2})^{1/2}\,\sin[2\alpha\,
{\rm arccos}(\tilde h)]\,/{2\alpha\tilde h}\,,\nonumber\\
&=&{{(1-\tilde h^2})}\,U_{2\alpha-1}(\tilde h)\,/{2\alpha\tilde h}
\label{Adef2}\,, \een where $U$ is the Chebyshew polynomials of
second kind; which explicit results, for $\alpha =2, 3$, are \ben
\wt A_2(\tilde h)&=&(1-\tilde h^2)(2\tilde h^2-1)\,,\label{A2def2}\\
\wt A_3(\tilde h)&=&\frac13(1-\tilde h^2)(1-2\tilde h^2) (3-4\tilde
h^2) \label{A3def2}\,. \een In this case, from eqs.~(\ref{wVtilde})
and (\ref{def2}), we have a family of models defined by the function
$\wt\omega_\alpha(\tilde h)$ and the potential $\wt V_\alpha(\tilde
h)$ written in polynomial form as \ben \wt \omega_\alpha(\tilde
h)&=&|\,[2\alpha\tilde h\,T_{2\alpha}(\tilde h)
+U_{2\alpha-1}(\tilde h)]/{4\alpha\tilde h^3}\, |\,,\label{wdef2}\\
\wt V_\alpha(\tilde h)&=&{(1-\tilde h^2)^2}\,U^2_{2\alpha-1}(\tilde
h)\,\wt \omega_\alpha(\tilde h)/{4\alpha^2}\,.\label{Vdef2} \een
Then, each value of the parameter $\alpha$ specifies a model of this
family. The explicit results for $\alpha = 2, 3$ are \ben
\wt\omega_2(\tilde h)&=&3-4\tilde h^2\,,\label{w2def2}\\
\wt V_2(\tilde h)&=&\tilde h^2(1-\tilde h^2)^2(1-2\tilde h^2)^2\,
\wt\omega_2(\tilde h)\,,\label{V2def2}\\
\wt\omega_3(\tilde h)&=&\frac13(19-64\tilde h^2+48\tilde h^4)\,,
\label{w3def2}\\
\wt V_3(\tilde h)&=&\frac19\tilde h^2(1-\tilde h^2)^2(1-2\tilde
h^2)^2 (3-4\tilde h^2)^2\,\wt\omega_3(\tilde h) \, . \nonumber
\\\label{V3def2} \een
\begin{figure}[ht]
\includegraphics[{height=3cm,width=07cm,angle=00}]{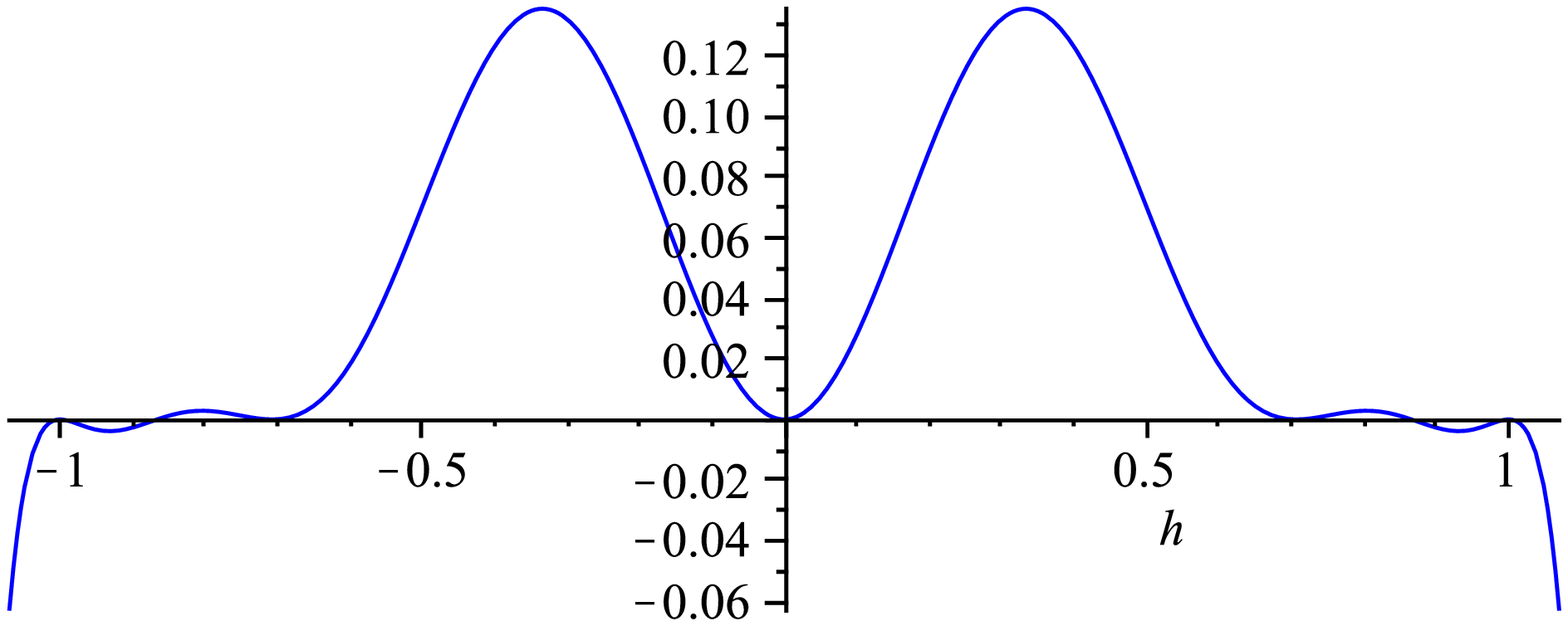}
\includegraphics[{height=2cm,width=07cm,angle=00}]{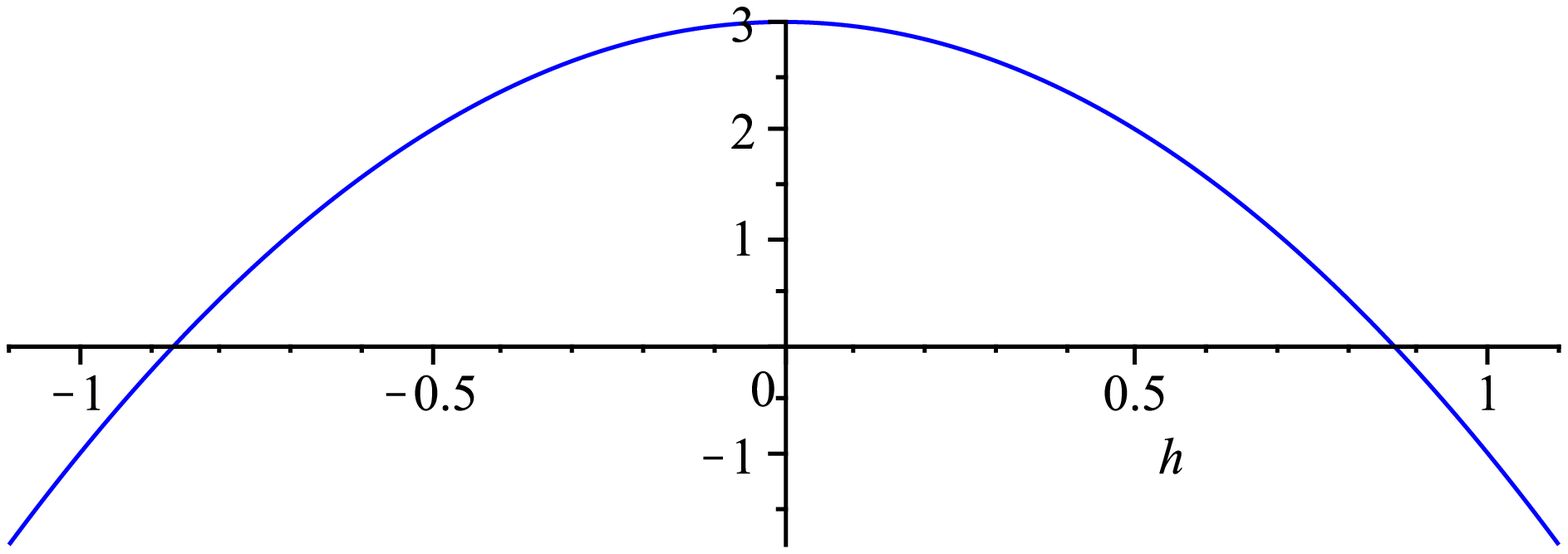}
\caption{The potential (\ref{V2def2}) (top panel) and the function
$w$ (\ref{w2def2}) (bottom panel), as function of $\tilde
h$.}\label{fig3a}
\end{figure}
For these models, from the inverse of deformation function
(\ref{def2}), we obtain the static Higgs field solutions in the form
\be\label{hdef2}
\tilde{h}^{(\pm)}(x)=\cos\left[({\arccos(h^{(\pm)}(x)) +
(m-1)\pi})/{\alpha}\right]\,, \ee where $h^{(\pm)}(x)$ is given by
eqs.~(\ref{hjac1}) and (\ref{hjac2}), and $m$ is an integer, which
generates distinct solutions only for $m=0, ..., \alpha-1$ .

\begin{figure}[ht]
\includegraphics[{height=2.5cm,width=4cm,angle=00}]{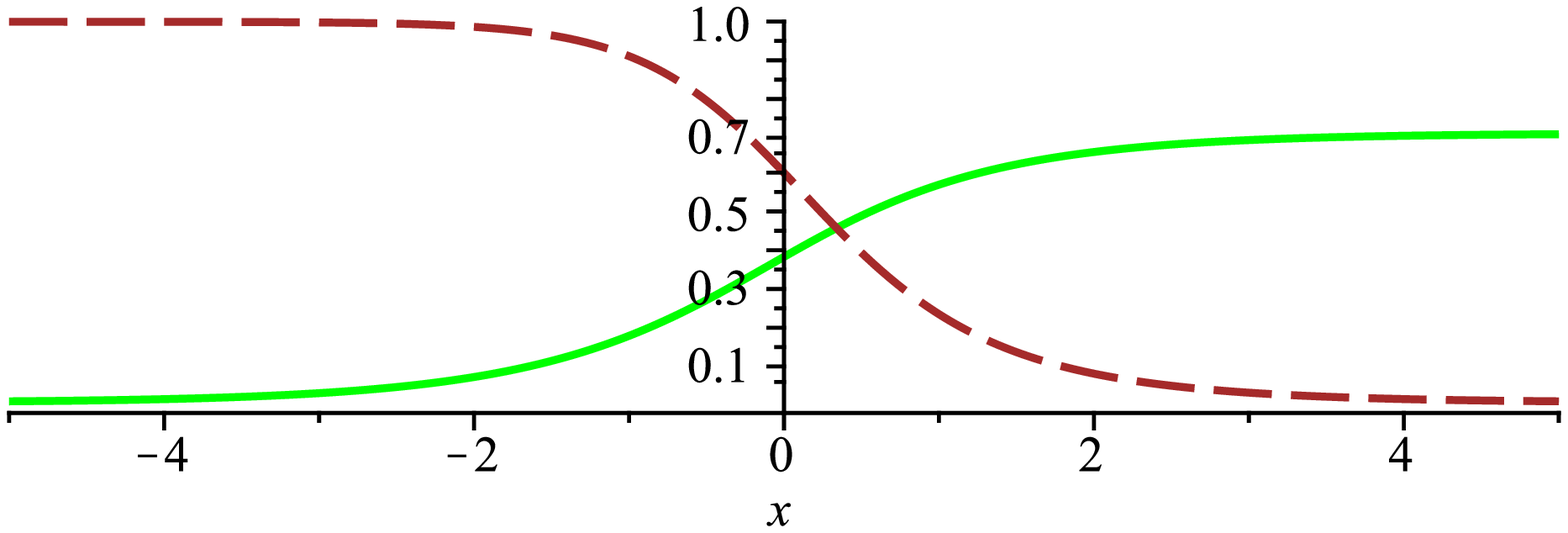}
\includegraphics[{height=2.5cm,width=4cm,angle=00}]{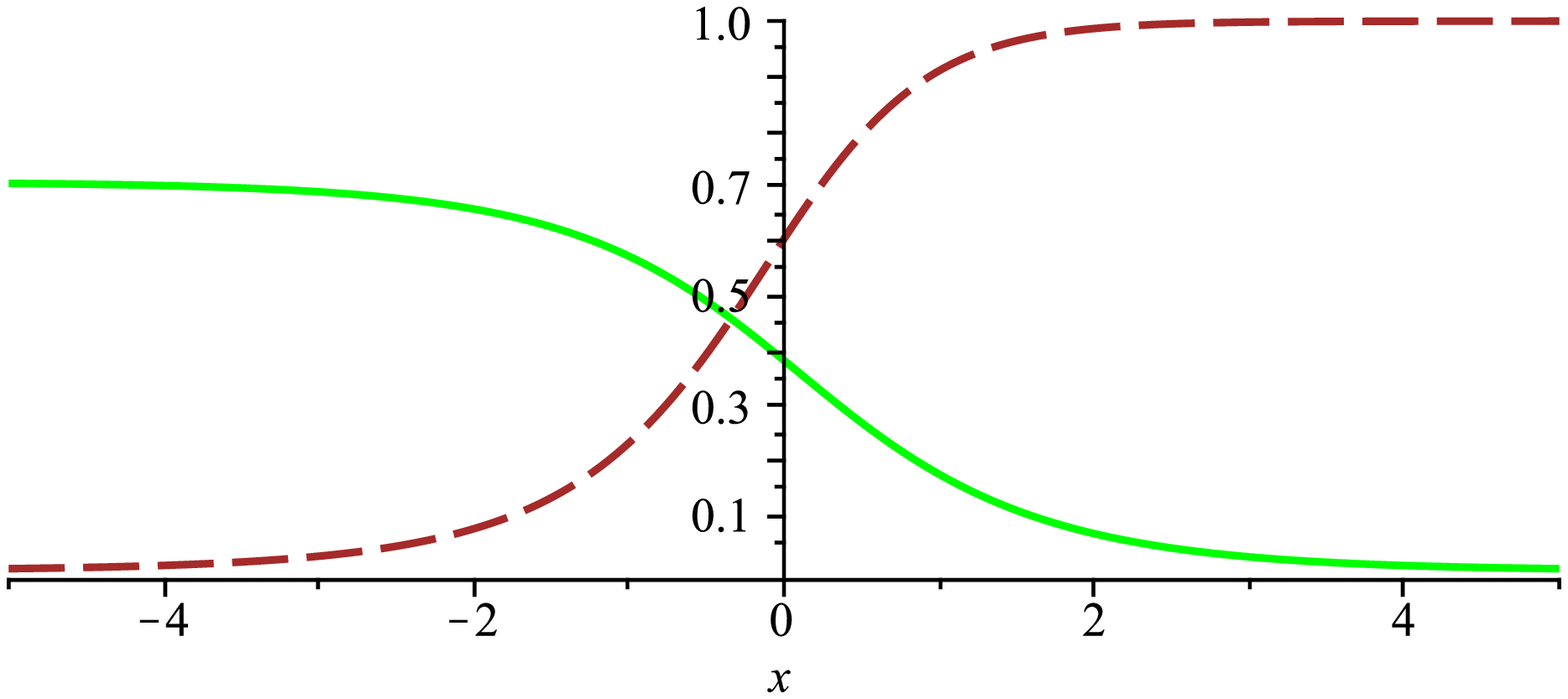}
\caption{The Higgs field (\ref{hdef2}) for $m=0$  (solid  line)  and
the  matching gauge field (\ref{A2def2}) (dashed line), for  $\tilde
h^{(+)}(x)$ and $\widetilde A^{(-)}(x)$, on the left, and for
$\tilde h^{(-)}(x)$ and $\widetilde A^{(+)}(x)$, on the
right.}\label{fig3c}
\end{figure}

\begin{figure}[ht]
\includegraphics[{height=2.5cm,width=4cm,angle=00}]{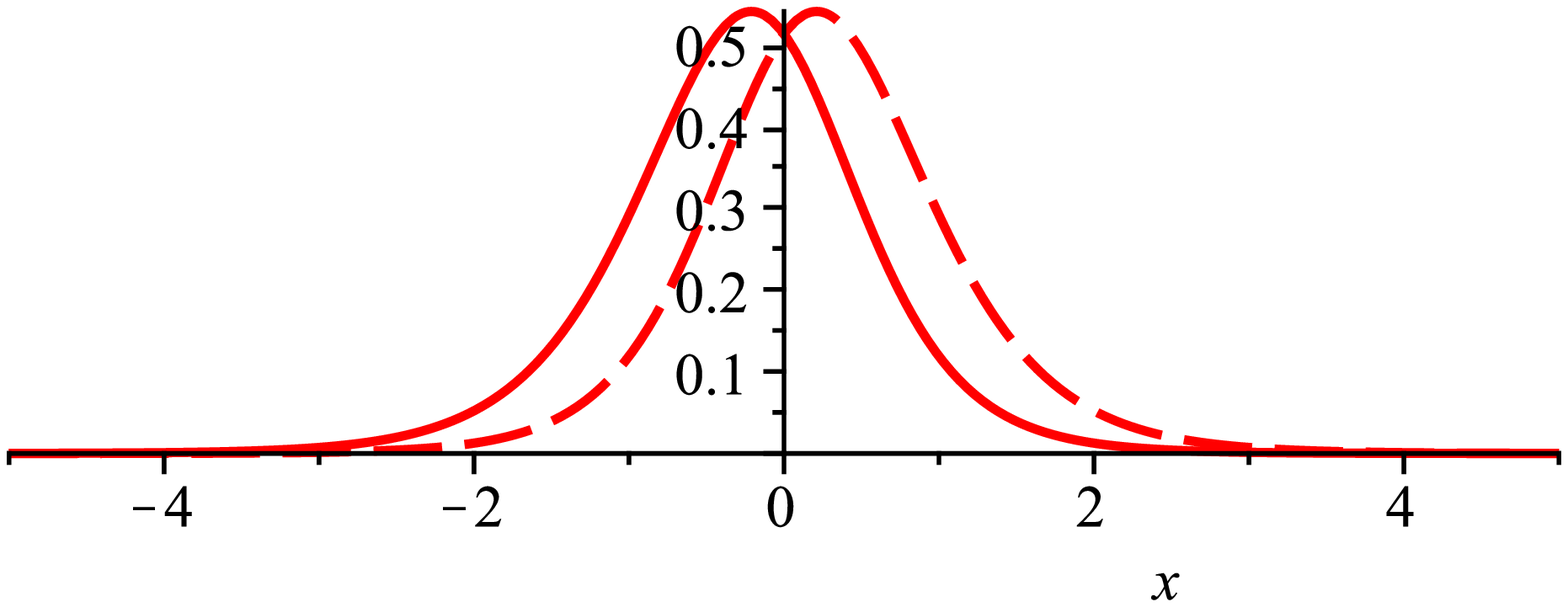}
\includegraphics[{height=2.5cm,width=4cm,angle=00}]{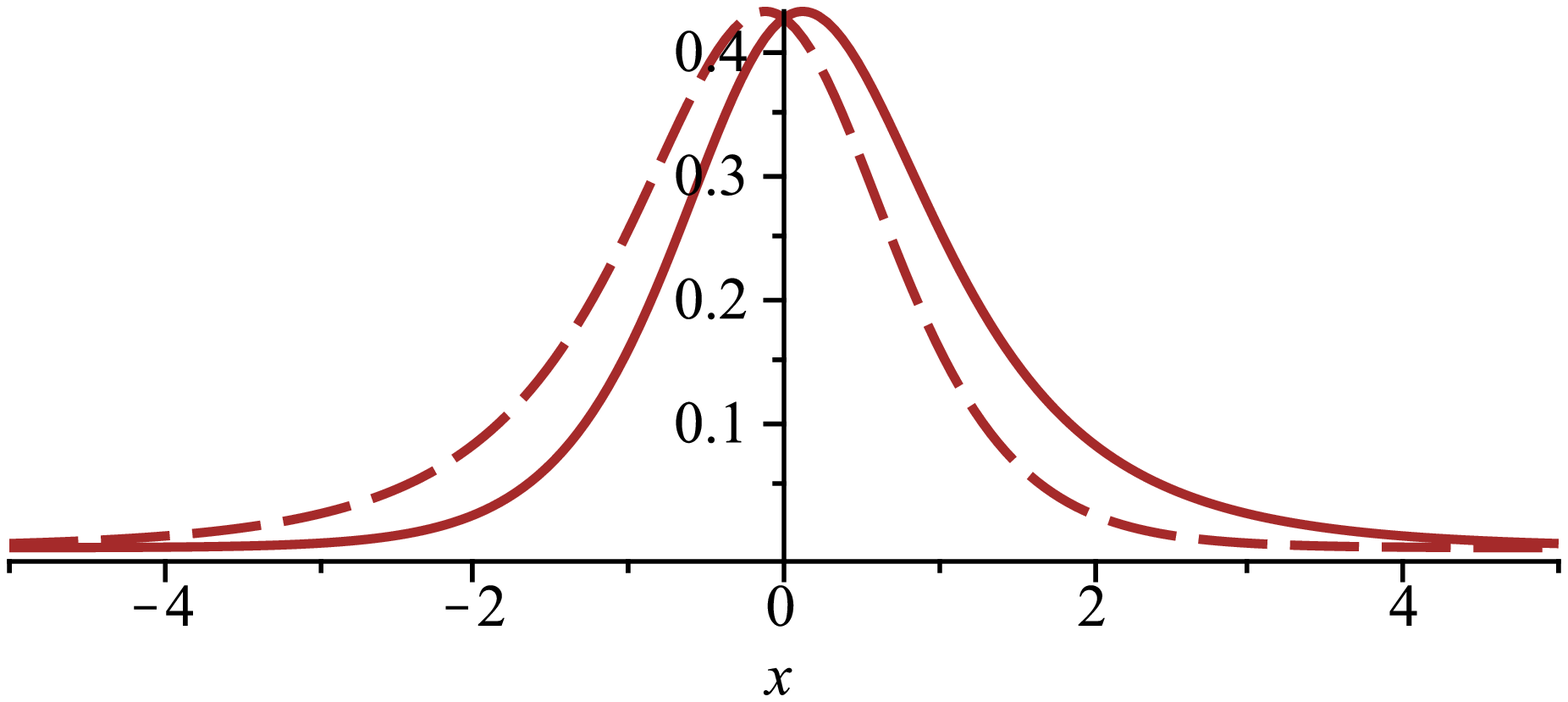}
\caption{The energy density (on the left) and the module of charge
density (on the right), for  solution $A^{(-)}(x)$ (solid line), and
for $A^{(+)}(x)$ (dashed-line).}\label{fig3e}
\end{figure}

Firstly, we examine the model for $\alpha=2$, defined by
eqs.~(\ref{w2def2}) and (\ref{V2def2}) displayed in
fig.~\ref{fig3a}. We see that, the potential is positive only for
$\tilde h\leq\sqrt{3/2}$. Then, there are two kind of  static
solutions for the Higgs field, one pair  kink/anti-kink  like
solution between $0\leq\tilde h\leq 1/\sqrt{2}$, and a lump-like
solution between $ 1/\sqrt{2}\leq\tilde h\leq\sqrt{3/2}$. In
Ref.~\cite{Santos11} is considered a model that presents a charged
lump-like solution. Here, the lump-like solution presents vanishing
charges and energy, hence we examine only the wall for $0\leq\tilde
h\leq 1/\sqrt{2}$. In fig.~\ref{fig3c}, we display the Higgs field
(\ref{hdef2}) and the gauge field (\ref{Adef2}) solutions.  These
walls have the same total energy and charges of the walls of the
standard JLW model, but with different spacial distribution of the
energy and charge densities, as shown in fig.~\ref{fig3e}.


\section{Ending Comments}

The examples presented above illustrate how the deformation method
may be used to generate many new generalized Abelian HCS models and
their defect solutions. This is achieved without requiring to
directly solve the nonlinear equations of motion of the new models.
The method also allows the construction of new defect solutions
controlling important features such as their height, width or the
topological character. Such results are of direct interest to
applications of domain walls in several contexts, such as
high-energy or condensed-matter physics.

\acknowledgements

Two of us (LL and JMCM) thank CAPES and CNPq (Brazilian agencies)
for financial support.

\end{document}